# Interplay of confinement and density on the heat transfer characteristics of nanoscale-confined gas

Reza Rabani [a], Ghassem Heidarinejad [a,*], Jens Harting [b,c], Ebrahim Shirani [d]

[a] Faculty of Mechanical Engineering, Tarbiat Modares University, 14115143, Tehran, Iran

[b] Helmholtz Institute Erlangen-Nürnberg for Renewable Energy (IEK-11), Forschungszentrum Jülich, Fürther Strasse 248, 90429 Nuremberg, Germany

[c] Department of Applied Physics, Eindhoven University of Technology, PO box 513, 5600MB Eindhoven, The Netherlands

[d] Department of Mechanical Engineering, Foolad Institute of Technology, 8491663763, Fooladshahr, Isfahan, Iran

[*] Corresponding author. Tel.: +98 21 82883361; fax: +98 21 88005040

E-mail addresses: reza.rabani@modares.ac.ir (R.Rabani); gheidari@modares.ac.ir, gheidari@alum.mit.edu (G. Heidarinejad*); j.harting@fz-juelich.de (J.Harting); eshirani@cc.iut.ac.ir, director@jafmonline.net (E.Shirani)

**Abstract:** The effect of changing the Knudsen number on the thermal properties of static argon gas within nanoscale confinement is investigated by three-dimensional molecular dynamics simulations. Utilizing thermalized channel walls, it is observed that regardless of the channel height and the gas density, the wall force field affects the density and temperature distributions within approximately 1 nm from each channel wall. As the gas density is increased for constant channel height, the relative effect of the wall force field on the motion of argon gas atoms and, consequently, the maximum normalized gas density near the walls is decreased. Therefore, for the same Knudsen number, the temperature jump for this case is higher than what is observed for the case in which the channel height changes at a constant gas density. The normalized effective thermal conductivity of the argon gas based on the heat flux that is obtained by implementation of the Irving–Kirkwood method reveals that the two cases give the same normalized effective thermal conductivity. For the constant density case, the total thermal resistance increases as the Knudsen number decreases while for the constant height case, it reduces considerably. Meanwhile, it is observed that regardless of the method used to change the Knudsen number, a considerable portion of the total thermal resistance refers to interfacial and wall force field thermal resistance even for near micrometer-sized channels. It is shown that while the local thermal conductivity in the near-wall region strongly depends on the gas density, the wall force field leads to a reduced local thermal conductivity as compared to the bulk region.

# 1. Introduction

With the rapid progress in fabricating and manufacturing of micro/nanoscale devices, understanding the fluid and heat transfer characteristics under nanoscale confinement becomes essential to improve the performance of the microfluidic device components and micro/nano electromechanical systems. Further examples which also deal with transport phenomena on these scales include the heat and mass transfer through carbon nanotubes [1,2] or the heat transfer between the head and disk in magnetic disk drives [3–5]. In such devices, gas experiences levels of rarefaction ranging from continuum behavior to the free molecular regime. The degree of rarefaction is characterised by the Knudsen number ($Kn = \lambda/H$) which is defined as the ratio of the mean free path, $\lambda$, to the characteristic length of the domain, $H$. The variation of rarefaction changes the transport mechanism from diffusive transport in the continuum regime to ballistic transport in the free molecular regime. Diffusive transport occurs when the characteristic length scale is larger than the gas mean free path. In contrast, ballistic transport is observed wherever the mean free path is larger than the characteristic length scale. Between these limiting conditions, the mean free path is of the same order as the characteristic length scale and a combination of diffusive and ballistic behavior is observed. This intermediate region is known as the transition regime. Considering a kinetic theory based method such as direct simulation Monte Carlo or the direct solution of the Boltzmann transport equation, gas flow and heat transfer in the transition regime have been studied extensively during the past decades [6–8].

In addition to non-equilibrium effects that arise due to the rarefaction, the interactions of the atoms or molecules with the surface and surface adsorption play an important role in the distribution of gas properties in nanoscale confinement. These phenomena are important within a couple of molecular diameters from the surface, i.e. at a scale that is comparable with the confinement. Therefore, at these scales, their effect on the distribution of hydrodynamic and thermal properties must not be neglected. Although kinetic theory can describe non-equilibrium effects precisely, the distribution of forces at the surface ("surface/wall force field") and surface adsorption phenomena can be modeled on the molecular-level by a technique such as molecular dynamics (MD). The effect of the surface force field on the adsorption of gas and liquid to the surface is mentioned in different studies [9–11]. By implementing purely attractive and repulsive wall models in three-dimensional molecular dynamics simulations of Poiseuille gas flow, density accumulation effects near the boundaries were observed and the difference in the velocity profile in comparison with continuum hydrodynamics was shown [12,13].

For a large domain size of the order of the mean free path in lateral and axial directions, several intermolecular collision times are required for time averaging and an excessive number of wall atoms is needed to model the surface atomistically. This leads to a high computational cost of MD simulations of gas flow in the high Knudsen number regime. To overcome this problem, the "smart wall molecular dynamics" (SWMD) algorithm was proposed and developed which is a cold wall model that reduces the number of wall atoms in MD simulations significantly. By using of the SWMD algorithm, nanoscale gas flow with 3-D molecular large surfaces was modeled. It was shown that the presence of a solid surface exerts a body force on the gas atoms. Due to the range of the wall force field, the gas distribution near the walls changes considerably. This region extends about 1nm from each wall and covers a significant volume of the channel [14].

Applying SWMD to stationary and shear driven nano-channel gas flows for various Knudsen numbers revealed that the presence of the wall force field leads to significant variations in the velocity- and the density profile from kinetic theory based calculation in the near wall region [15,16]. A comprehensive study on transport phenomena in bulk and near-wall regions implied that the gas–wall interaction parameter and the Knudsen number determine the transport characteristics [16–18]. In another study on force-driven nano-channel gas flows for different Knudsen regimes, a new dimensionless parameter $\beta$ was defined as the ratio of the wall force penetration length to the channel height $(\beta = l_f/H)$. It was shown that for a finite value of this parameter, the near-wall region covers larger portions of the flow domain which leads to a deviation from kinetic theory's prediction of mass transport [19]. As an inner layer scaling, $y^* = y/\sigma$ was introduced based on the molecular diameter $(\sigma)$ recently. It was shown that a universal behavior as a function of the local Knudsen number and the gas–wall interaction parameters exists for velocity profiles within $(y^* < 3)$. Furthermore, a procedure that can correct kinetic-theory-based mass flow rate predictions for various nano-channel gas flows was also presented [20].

The preceding review reveals that several aspects of the effect of the wall force field on the distribution of the gas properties within nanochannels were studied extensively by the use of SWMD. Due to the cold wall nature of SWMD, all these simulations were done under isothermal conditions of the gas and the surface by applying a thermostat to the gas medium. Actually, in many applications, there is considerable heat transfer between the gas and walls that must not be neglected in the simulations. The heat flux might arise due to the difference between the temperature of the walls and the gas or even for the isothermal condition between the walls and the gas while the viscous heating is considerable. Since it was shown that a steep velocity gradient exists in the gas flow in the nanoconfined medium due to the presence of the wall force field [14,16,17,19], it is expected that the viscous heating is considerable in the nanoconfined medium. Therefore, the effect of the wall force field on the thermal behavior of the gas in nanoconfinement should also be considered in the MD simulation. The existing literature reveals that the heat transfer characteristics determined by the heat flux, the temperature distribution, thermal conductivity, etc., of the confined liquid has been studied extensively utilizing appropriate wall/gas interaction potentials [21–31]. However, the distribution of thermally related properties in nanoscale confined gases are not clear yet.

The objective of this paper is to investigate the effect of the wall force field on thermal properties such as the temperature profile, heat flux, thermal conductivity and thermal resistance of a simple gas confined within a nanoscale channel. MD simulations for the monoatomic noble gas atoms of argon at various density and channel height are conducted. Changing the mean free path and the length scale of the domain result in the variation of the Knudsen number by two different mechanisms. This reveals a different impact of the wall force field on the heat transfer. In the first set of simulations, the constant height case (denoted as H~), the height of the channel is kept constant and the mean free path of gas atoms is varied by changing the density of the gas in a way that the modified Knudsen number, $k = (\sqrt{\pi}/2)\, Kn$, varies from the early transition to the free molecular regime. It is expected that in this case, the combined effect of wall force field and Knudsen number on the distribution of the gas properties is observed. In the second set of simulations, the constant density case (denoted as ρ~), the density of the gas is kept constant and the height of the channel is increased in a way that $k$ varies within the same range as in the previous set of simulations. Since increasing the channel height decreases the relevant importance of the wall force field on the gas properties, it is anticipated that only the effect of changing the Knudsen number is observed in this case. By comparing these two cases, we are able to calculate the relative importance of the wall force field on thermal properties of the gas. To the knowledge of the authors, this work presents the effect of the wall force field on the thermal properties of nanochannel-confined gas for the first time in the literature.

The remainder of this article organized as follows: In section 2, we describe the MD algorithm and simulation parameters as well as the computation of the heat flux. Section 3 presents the validity of the solution by comparing with previously published data [15]. In Section 4, we present the effect of the wall force field for two different mechanisms of variation of the Knudsen number in the transition regime by comparing the gas density, temperature, heat flux, thermal conductivity and thermal resistance distribution. Finally, we conclude in Section 5.

## 2. Three-dimensional MD simulation

We use the molecular dynamics code LAMMPS (Large-Scale Atomic/Molecular Massively Parallel Simulator) from Sandia National Laboratories to simulate a confined argon gas between two parallel plates that are a distance $H$ apart [32]. In the streamwise ($W$) and lateral ($L$) directions, periodic boundary conditions are applied. In order to obtain a solution which is independent of the domain size, the computational domain extends at least for one mean free path (54 nm) in the periodic direction [14]. The domain size for each case is specified in Table 1 and the exact number of argon atoms is chosen according to the previous study [16]. To model the van der Waals interactions between different atoms, a truncated (6–12) Lennard–Jones (L–J) potential is used,

$$\emptyset_{truncated}(r_{ij}) = \begin{cases} 4\varepsilon\left[\left(\frac{\sigma}{r_{ij}}\right)^{12} - \left(\frac{\sigma}{r_{ij}}\right)^{6}\right] - \emptyset(r_C) & r \leq r_C \\ 0 & r > r_C \end{cases}, \quad (1)$$

where $r_{ij}$ is the interatomic distance, $r_c$ is the cutoff distance and $\emptyset(r_C)$ is the value of the interatomic potential at $r = r_C$. The mass of an argon atom is $m = 6.63 \times 10^{-26}$ kg, the argon atomic diameter is $\sigma = 0.3405$ nm and the depth of the potential well for argon is $\varepsilon = 119.8 \times k_b$, where the Boltzmann constant is $k_b = 1.3806 \times 10^{-23}$ J/K. Considering the fact that the L-J potential is negligible at a large molecular distance, it is common to set the cutoff distance to $r_C = 1.08$ nm for a dilute gas, whereas $r_C = 2.7$ nm is considered for dense gases [15,33]. For the simulation of the walls, the same molecular mass and diameter as for the argon gas are considered ($m_{\text{wall}} = m_{\text{Ar}}$, $\sigma_{\text{wall}} = \sigma_{\text{Ar}}$). Furthermore, it is assumed that the potential strength of the gas-wall interactions is equal to the potential strength of the gas-gas interaction ($\varepsilon_{\text{wall}-\text{Ar}} = \varepsilon_{\text{Ar}-\text{Ar}}$) [14–16,18,19]. Considering the cutoff radius, two layers of FCC (face-centered cubic) aligned particles are used to model the wall [14–16].

The motion of the gas and wall atoms is determined by Newton's second law using a velocity Verlet algorithm for the time integration [33]. In order to control the wall temperature, the so-called "interactive thermal wall model" (ITWM) is used [34]. In this method, the wall particles are fixed onto their initial lattice sites by springs and oscillate around their equilibrium position to exchange momentum and energy with the fluid particles through intermolecular interactions and collisions. On each layer of wall atoms, a velocity-scaling thermostat is applied in order to assure a uniform temperature distribution in the walls. In this way, there is no need to apply any thermostat on the gas and the heat is properly transferred to/from the gas through the walls [34,35]. In the present study, the value of $k_s = 500\varepsilon\sigma^{-2}$ is used as the wall stiffness which determines the strength of the bonds between the wall particles [36].

The initial temperature is set to 298 K for argon and wall atoms in all simulations. Simulations are started from the Maxwell–Boltzmann velocity distribution for all atoms at this temperature. To reach thermal equilibrium, we let the initial particle distribution evolve for $5 \times 10^6$ time steps. This initial procedure ensures that gas and wall atoms attain their equilibrium by interacting with each other. Afterwards, in order to induce a heat flux in our system, a different temperature is applied at the top ($T_H = 308$ K) and the bottom wall ($T_C = 288$ K). Depending on the implemented temperature differences between the walls and the Knudsen number, $n_{ssc}$ time steps are performed to attain the steady state with the new heat flux conditions. Then, $n_{ave}$ time steps are done for averaging microscopic quantities to obtain macroscopic properties of the gas. The exact values for $n_{ssc}$, $n_{avr}$ and other simulation parameters are listed in Table 1. Longer time averaging is also performed in each case to confirm convergence of macroscopic quantities such as the density profile, the temperature profile and the heat flux. In all simulations, the NVE ensemble is applied to all atoms in the domain. The channel height is determined from the centerlines of the first layer of wall atoms of the top and bottom surfaces. The computational domain is divided into bins to obtain averaged quantities. To assure that the bin size is appropriate, we compare the averaged quantities to values obtained from bins of four different sizes. While for the bulk region larger bins would suffice, we choose a bin size of approximately $\sigma/10$ to resolve the features of the temperature profiles in the near wall region.

Different temperatures for walls will lead to a thermal gradient in the perpendicular direction and this gradient generates a heat flux between the walls. The heat flux vector is determined from the Irving–Kirkwood (I–K) expression as [37,38]:

$$J_l = \frac{1}{\text{Vol}} \langle \sum_i V_l^i E_{tot}^i + \frac{1}{2} \sum_{i,j} r_l^{ij} (f^{ij} \cdot V^i) \rangle, \qquad (2)$$

$$E_{tot}^i = \frac{1}{2} m^i \left( (V_x^i)^2 + (V_y^i)^2 + (V_z^i)^2 \right) + \emptyset^i, \qquad (3)$$

where the summation is performed over all argon gas atoms. Considering $l$ as the axes of the Cartesian coordinate system, $V_l^i$ is the velocity component of particle $i$ in the $l$-direction. Furthermore, $E_{tot}^i$ is the total and $\emptyset^i$ is the potential energy of particle $i$ which are calculated using Equations 3 and 1, respectively, and $r_l^{ij}$ is the distance vector between particle $i$ and $j$. In addition to that, $f^{ij}$ is the vector of intermolecular force exerted on particle $i$ by particle $j$ and $V^i$ is the velocity vector. Furthermore, it should be mentioned that in some figures, only a few averaging points are shown in order to present the curves in that figure more clearly.

## 3. Validation

The validity of the results is investigated by comparing the temperature and density profiles with the data of Barisik and Beskok [15]. Argon at $\rho = 1.896$ kg/m$^3$ is considered in a domain of size $54 \times 5.4 \times 54$ nm and under isothermal conditions with walls at a temperature of 298 K. This temperature is applied on the walls and the gas and we let the initial particle distribution evolve for 10 ns to reach thermal equilibrium. For the time averaging of microscopic quantities to obtain macroscopic quantities, at least 20 ns are considered. The normalized density and temperature distributions of gas in ITWM that is used in this study are compared to the value of the SWMD in Fig. 1(a) and Fig. 1(b), respectively. In this manuscript, all densities are normalized to their corresponding value in the middle of the channel. Both Figures show a very good agreement of our work with the literature data. It is important to stress that in the SWMD, the Nose-Hoover thermostat is applied on the argon gas to keep the gas temperature at the desired value, while in the ITWM, the thermostat is applied on the walls. Considering Figs. 1(a) and 1(b), the only mismatch between these two methods refers to the temperature distribution in Fig. 1b where a small difference is observed in the wall region. This is due to the presence of the rigid walls in the SWMD which are by definition athermal so the temperature is not defined there. In contrast, in our simulation the ITWM method produces the specified temperature at the wall which in here is identical to the temperature of the gas.

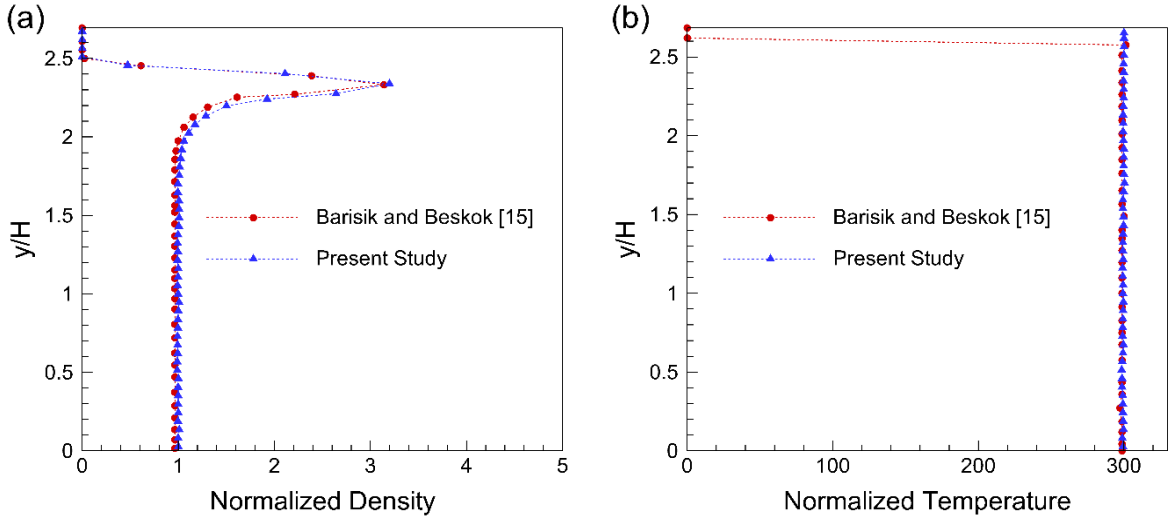

**Fig. 1.** Normalized density **(a)** and temperature **(b)** distribution of argon gas at 298 K for ITWM and SWMD

## 4. Results and discussion

In order to investigate the effect of the wall force field on the distribution of thermal properties for various Knudsen numbers in the transition regime, two different mechanisms of changing the Knudsen number are considered (see Table 1 for the simulation parameters). In order to have the desired Knudsen number for each case, the corresponding number of argon gas atoms is chosen following Barisik and Beskok [16]. In the constant density case, the density of the argon gas is kept constant (at approximately $\rho = 1.896$ kg/m$^3$) and the height of the channel is varied in such a way that the Knudsen number covers the whole transition regime. Figure 2 shows the schematic variation of the height and the corresponding Knudsen number for various cases which are studied here. Since the wall force field extends approximately within 1 nm from each wall and the height of the channels increases as the Knudsen number is decreased, it is expected that the effect of the wall force field decreases gradually. In the constant height case, the height of the channel is kept constant (as $H = 5.4$ nm) and the density is changed such that the Knudsen number is the same as in the previous case. In Fig. 3, the schematic variation of the height and the corresponding modified Knudsen number are shown. Since the channel height is constant at 5.4 nm and the wall force field extends approximately in 1 nm from each wall, approximately 40% of channels height is affected by this force. Therefore, it is expected that in this set of simulations, the wall force field affects the whole range of Knudsen numbers and the properties of the argon gas are changed based on that.

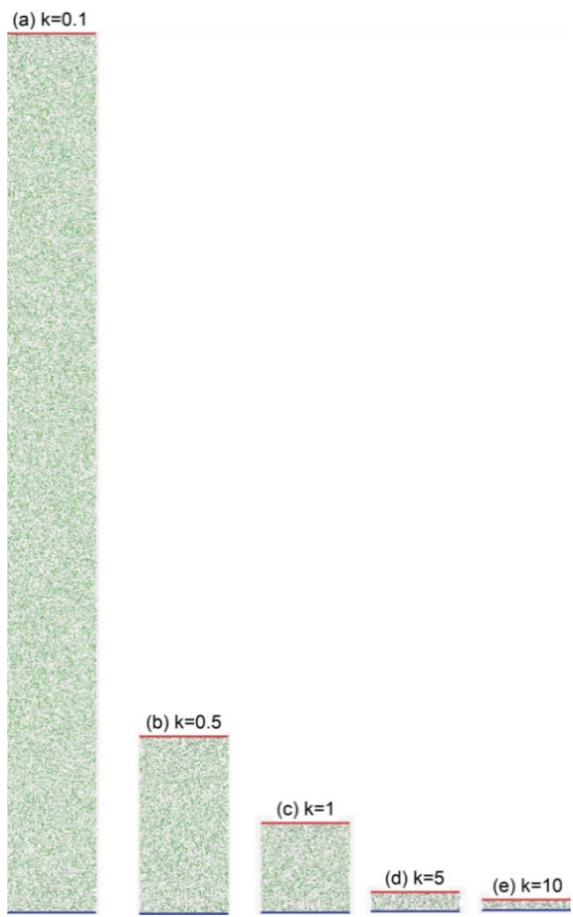

**Fig. 2.** Variation of the channel height for different modified Knudsen numbers in the transition regime (case H~)

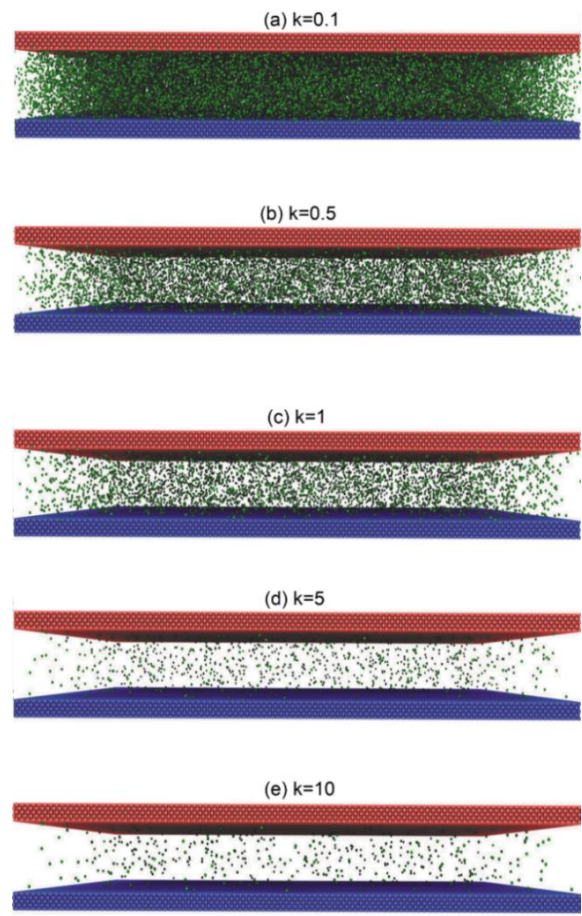

**Fig. 3.** Variation of the density for different modified Knudsen numbers in the transition regime (case ρ~)

It should be noticed that a comparison between the bulk gas density of the present work (Table 1) and Barisik and Beskok [16] reveals that with the same initial number of gas atoms, the ITWM wall model leads to a smaller value for the bulk gas density than the SWMD wall model (approximately 5% reduction is observed). This discrepancy is referred to the nature of the ITWM wall model where the finite spring constant between the wall atoms lets the gas atoms to become closer to the walls so that the bulk gas density decreases slightly. Therefore, in order to have the desired bulk gas density in each case, the number of gas atoms is increased slightly in comparison with Barisik and Beskok [16] according to Table 1.

The normalized density and temperature distributions for varied channel height at a constant density in the transition regime are shown in Figs. 4(a) and (b), respectively. Both figures clearly show that as the height of the channel increases, the wall force field affects a smaller portion of the channel's height. For all cases in Fig. 4(a), it is observed that the density is constant in the bulk region of the gas and as the channel height is increased, the ratio of the channel height that is affected by the wall force field goes to zero (It is defined as wall force field penetration depth divided by the channel height). Fig. 4(b) shows that in the bulk portion of the channel where the effect of the wall force field is negligible, by decreasing the Knudsen number, the temperature profile changes from approximately a constant value to a linear distribution. It should be mentioned that in order to obtain clear profiles in Figs. 4(a) and (b), the averaged value of the coarse bin is depicted for $H = 54, 108$ and $540$ nm.

**Table 1** MD simulation details for the constant density and constant height case

| $k$ | $W \times H \times L$ (nm) | # argon atoms | $\rho$ (kg/m$^3$) | $r_C$ (nm) | $n_{ssc}$ | $n_{ave}$ | $k_C$ (mW/mK) |
|---|---|---|---|---|---|---|---|
| 0.1 ($H\sim$) | $54 \times 540 \times 54$ | 45300 | 1.896 | 1.08 | $20 \times 10^6$ | $80 \times 10^6$ | 17.85 |
| 0.1 ($\rho\sim$) | $54 \times 5.4 \times 54$ | 46200 | 189.6 | 2.7 | $1 \times 10^6$ | $15 \times 10^6$ | 22.95 |
| 0.5 ($H\sim$) | $54 \times 108 \times 54$ | 9100 | 1.896 | 1.08 | $20 \times 10^6$ | $80 \times 10^6$ | 17.85 |
| 0.5 ($\rho\sim$) | $54 \times 5.4 \times 54$ | 9400 | 37.92 | 2.7 | $1 \times 10^6$ | $20 \times 10^6$ | 18.88 |
| 1 ($H\sim$) | $54 \times 54 \times 54$ | 4550 | 1.896 | 1.08 | $20 \times 10^6$ | $80 \times 10^6$ | 17.85 |
| 1 ($\rho\sim$) | $54 \times 5.4 \times 54$ | 4700 | 18.96 | 2.7 | $5 \times 10^6$ | $25 \times 10^6$ | 18.33 |
| 5 ($H\sim$) | $54 \times 10.8 \times 54$ | 920 | 1.896 | 1.08 | $20 \times 10^6$ | $80 \times 10^6$ | 17.85 |
| 5 ($\rho\sim$) | $54 \times 5.4 \times 54$ | 940 | 3.792 | 1.08 | $5 \times 10^6$ | $40 \times 10^6$ | 17.91 |
| 10 | $54 \times 5.4 \times 54$ | 460 | 1.896 | 1.08 | $20 \times 10^6$ | $80 \times 10^6$ | 17.85 |

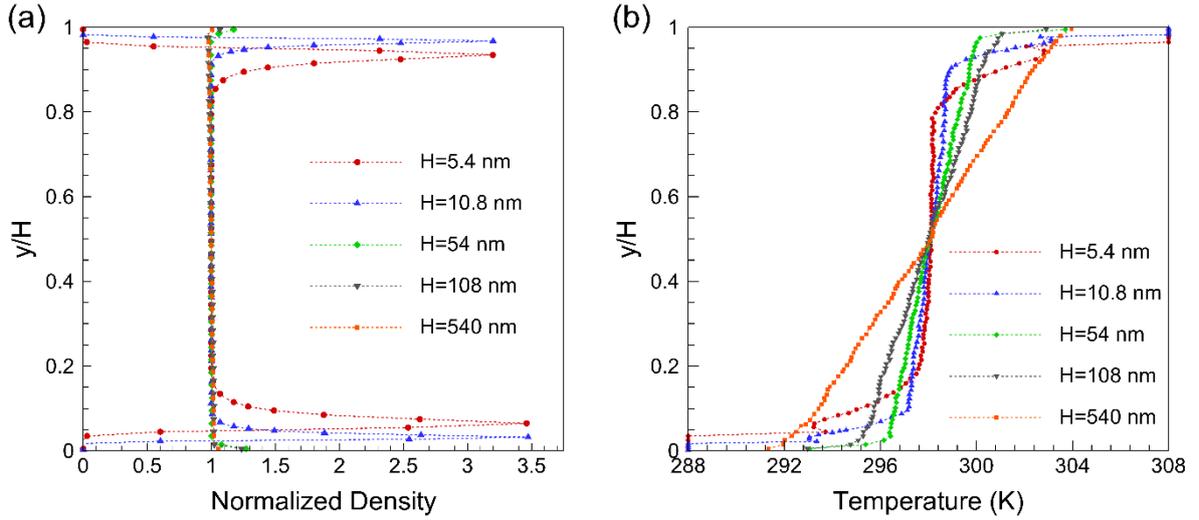

**Fig. 4.** Normalized density **(a)** and temperature **(b)** profile along the channel height for the H~ case

The variation of the normalized density at a distance of 2 nm from the cold and the hot wall is depicted in Figs. 5(a) and (b), respectively. Since the density is constant for all cases, the normalized density near the wall region shows similar behavior regardless of the Knudsen number. It is obvious that the wall force field increases the residence time of argon gas atoms which results in an increased density near the wall region [14–16]. It should be mentioned that the exact amount of the increment in the residence time of the gas atoms has not been studied yet. This quantity can be calculated with the same method which was applied by Sofos et al. [39]. However, its overall effect on the gas distribution is important in our analysis which is shown clearly in the Fig. 5.

Considering Fig. 5, it is observed that for all Knudsen numbers, the maximum value of the normalized density near the cold wall is about 3.54 times greater than the bulk density which is reduced to 3.19 for the hot wall. This difference refers to the fact that the cold wall absorbs the energy of impinging argon atoms. Therefore, it takes more time to escape from the wall force field region and the increase in the residence time results in a higher accumulation of argon atoms near the cold wall. In contrast, the hot wall increases the energy of impinging argon atoms resulting in reduced residence times. It should be emphasized that for all cases in this manuscript, the temperature difference is 20 K and this density difference is expected to become higher for larger temperature differences between the walls.

Figures 6(a) and (b) show the variation of the temperature distribution within 2 nm distance from the cold and hot walls, respectively. It can be seen that the temperature profile is affected by the wall force field for all Knudsen numbers. Since the gas density and temperature are the same at all heights, the different behavior as depicted by the temperature profile is only due to the difference between the Knudsen numbers. Considering the temperature jump as the difference between the maximum temperature of argon gas near each wall and the wall temperature, Fig. 6(a) shows that by increasing the channel height from $H = 5.4$ nm to $H = 540$ nm (corresponding to a change of $k = 10$ to $k = 0.1$) the temperature jump decreases from 5.2 K to 1.9 K. The same behavior is observed for the hot wall. Here, the temperature jump decreases from 5.46 K to 2 K while the channel height is increased from $H = 5.4$ nm to $H = 540$ nm.

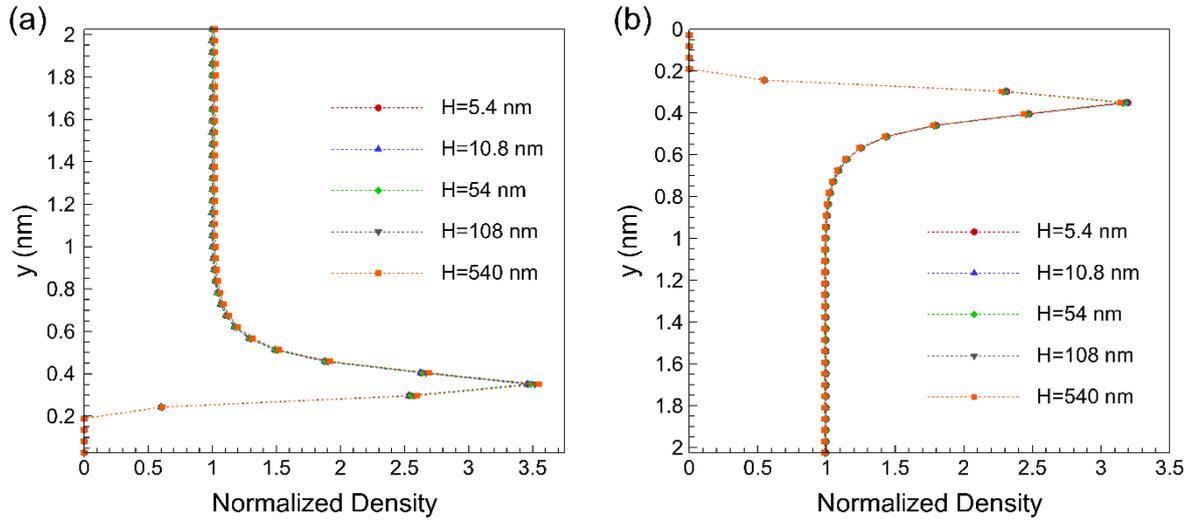

**Fig. 5.** Normalized density variation along the channel height within 2 nm from the cold **(a)** and the hot wall **(b)** for the H~ case

The normalized density and temperature distributions for varied gas density in the constant height case are shown in Figs. 7(a) and (b), respectively. Unlike the previous case, the argon gas status changes from rarefied for $k = 10$ to dense for $k = 0.1$. Fig. 7(a) clearly shows that in the bulk region, the normalized gas density coincides for all cases regardless of the argon gas density. It should be noticed that the density distribution for $k = 0.1$ shows the onset of density layering which is expected since the gas is dense. Furthermore, as the Knudsen number changes from $k = 10$ to $k = 0.1$ the maximum value of normalized density near the cold wall varies from 3.46 to 3.22 while for the hot wall it varies from 3.22 to 2.92. This reduction in the normalized density near the walls refers to a gas density increase.

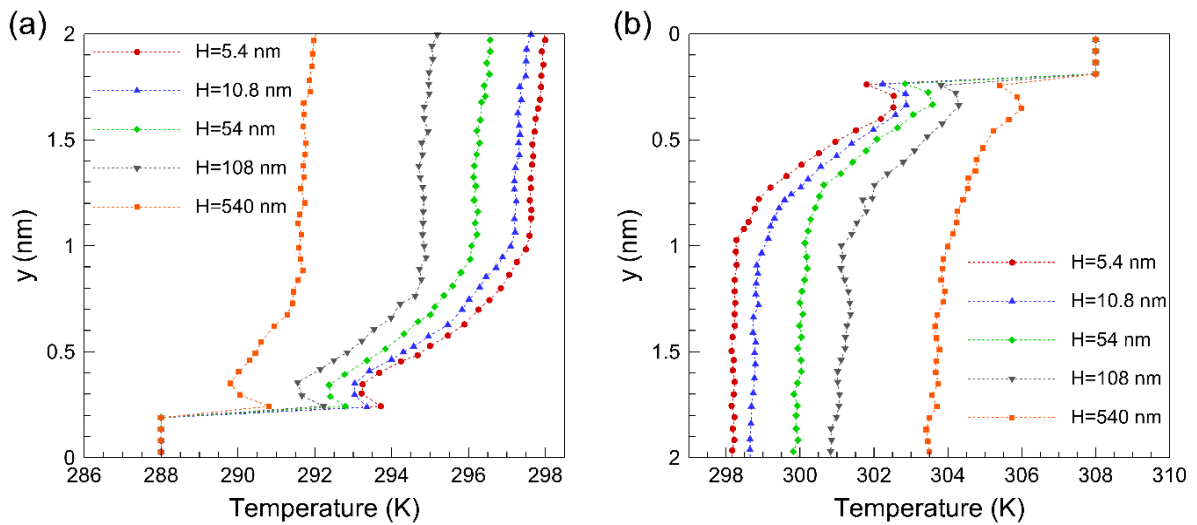

**Fig. 6.** Temperature distribution along the channel height within 2 nm from the cold **(a)** and the hot wall **(b)** for the H~ case

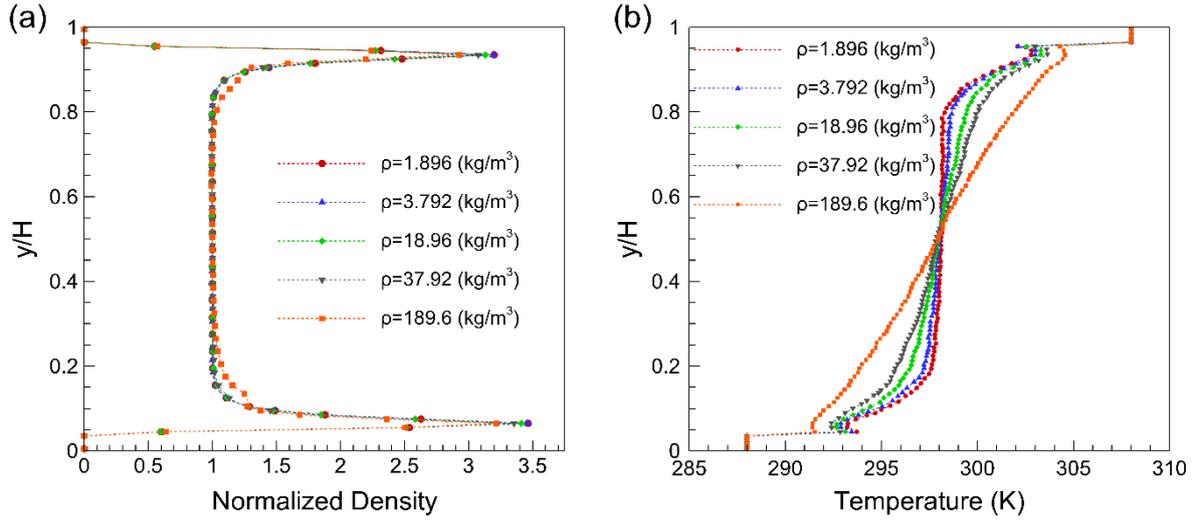

**Fig. 7.** Normalized density **(a)** and temperature **(b)** profile along the channel height for the ρ~ case

According to Fig. 3 for $k = 10$, when an argon atom reaches the wall, the main force that affects its motion is the wall force field. Since the gas is rarefied, the interaction between the argon atoms is negligible in comparison to the wall force field. As the density increases to $k = 0.1$, the argon atoms get closer to each other. Therefore, in the near wall region, the motion of the argon gas atoms is affected by the wall force field and forces induced by other argon atoms simultaneously. Therefore, the residence time in the wall force field region changes which affects the accumulation of argon atoms near the walls. The temperature distribution in Fig. 7(b) reveals that similar to the constant density case, reducing the Knudsen number changes the temperature profile in the bulk region from an approximately constant value to a linear variation between the walls.

The variation of the temperature distribution for various Knudsen numbers within 2 nm distance from the cold and the hot wall is shown in Fig. 8(a) and (b), respectively. Fig. 8(a) clearly shows that increasing the argon gas density (corresponding to a change of $k = 10$ to $k = 0.1$) decreases the temperature jump from 5.2 K to 3.4 K. The same behavior is observed for the hot wall where the temperature jump decreases from 5.46 K to 3.44 K while the argon gas density is increased from $1.896 \text{ kg/m}^3$ to $189.6 \text{ kg/m}^3$.

It should be mentioned that for a dense gas ($k = 0.1$), the effective range of the wall force field on the temperature profile is reduced to approximately 0.5 nm while for a rarefied gas ($k = 10$) it is about 1 nm. As mentioned before, this difference refers to the fact that for the constant height case, many argon atoms are in the vicinity of the wall and the wall force field is not the dominant force in the near wall region anymore. Here, the interaction between gas atoms affect this region, too.

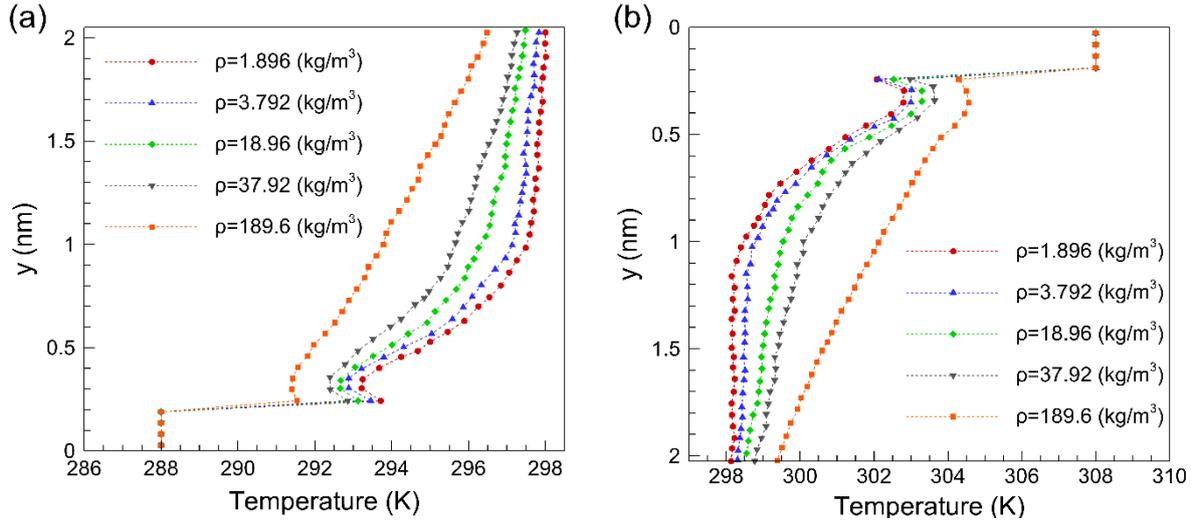

**Fig. 8.** Temperature distribution along the channel height within 2 nm from the cold **(a)** and the hot wall **(b)** for the ρ~ case

In Figs. 9(a) and (b), we compare the normalized density and temperature profile for $k = 0.1$ and $k = 1$ for the two different methods of changing the Knudsen number. As it is clear that in the bulk region where the wall force field is negligible, the normalized density and the temperature profile coincide with each other. It is obvious that by moving toward the wall, the ratio of the channel height that is affected by the wall force field is different for the same Knudsen number in both cases. Furthermore, the onset of density layering is observed for $k = 0.1$ when the density has been changed for the constant height case while for the constant density case, it is not observed anymore. Considering these observations, we find that the dynamic similarity assumption breaks down in the wall force field region for the normalized density and temperature profile between two different mechanisms of changing the Knudsen number while it is still valid in the bulk region of the gas. This conclusion is in agreement with a previous observation on the dynamic similarity assumption which was based on the comparison of the velocity profiles obtained from the constant density and constant height cases [19].

Using the Irving–Kirkwood (I–K) expression, the heat flux is depicted in Fig. 10(a) for both cases in the transition regime. It shows that in the constant density case when the channels height increases from $H = 5.4$ nm to $H = 540$ nm, the heat flux reduces from $0.765$ MW/m$^2$ to $0.35$ MW/m$^2$. This is due to the fact that when the temperature gradient decreases by increasing the channel height, a reduction in heat flux is expected. Meanwhile, for the constant height case as the gas density increased from $\rho = 1.896$ kg/m$^3$ to $\rho = 189.6$ kg/m$^3$, the heat flux through the gas increases from $0.765$ MW/m$^2$ to $45.5$ MW/m$^2$. In such a situation, the temperature gradient is constant and since the gas density gradually increases, an increase in heat flux is expected. The effective thermal conductivity in the constant density and constant height case is depicted in Fig. 10(b). Considering the fact that the Knudsen number is high, Fourier's law is not valid anymore to determine the thermal conductivity of the argon gas. Therefore, we use an "effective thermal conductivity" $K_{eff} = J \, \Delta T/H$ [40]. As it is expected by reducing the Knudsen number, the effective thermal conductivity for both, the constant density and the constant height cases, increases.

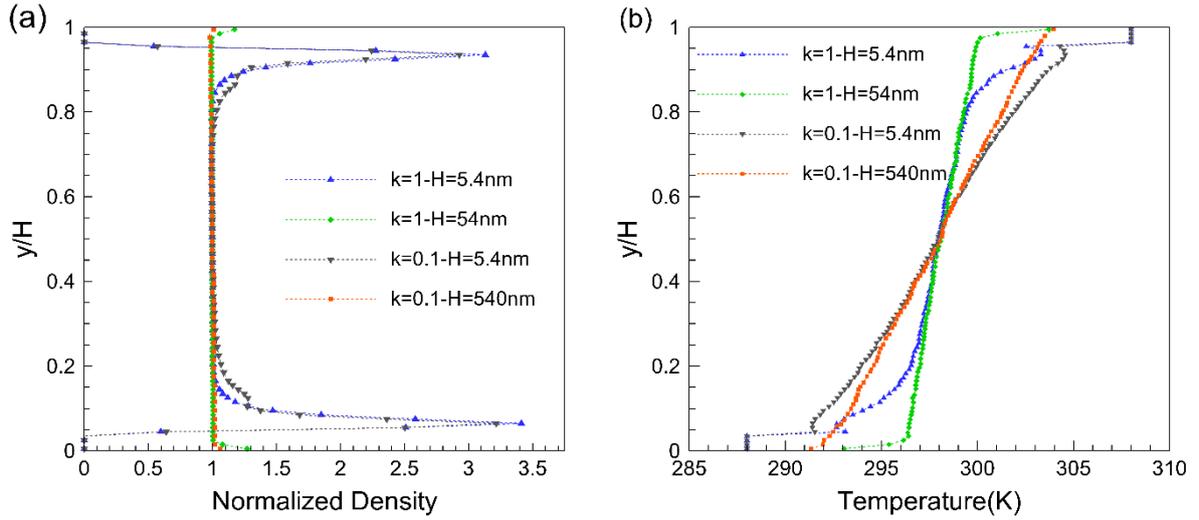

**Fig. 9.** Comparison between normalized density **(a)** and temperature **(b)** distribution along the channel height for different methods of changing the Knudsen number (comparison between the cases ρ~ and H~)

Using the thermal conductivity of argon gas at the desired density and temperature $k_C$ [41] from Table 1, the effective thermal conductivity is nondimensionalized and plotted in Fig. 10(c). This Figure clearly shows that independent of the method of changing the Knudsen number, the normalized effective thermal conductivity is identical in both cases. In order to use the thermal resistance definition $R_i = \Delta T_i/J$ to calculate the total thermal resistance of the system, the temperature profile is divided into five separate parts where in each part the temperature distribution is approximately linearly as it is shown in Fig. 11(a). According to this notation, $R_{AB}$ is the interfacial thermal resistance of the cold wall and $R_{EF}$ is the interfacial thermal resistance of the hot wall. The total thermal resistance of the argon gas is defined as

$$R_{Tot} = R_{AB} + R_{BC} + R_{CD} + R_{DE} + R_{EF}. \qquad (4)$$

The value of the thermal resistance for each part of the channel height is calculated and presented in Table 2. Furthermore, the total interfacial thermal resistance between the argon gas and the walls is defined as the summation of the interfacial thermal resistance of both walls as:

$$R_{Interfacial} = R_{AB} + R_{EF}. \qquad (5)$$

In addition, as it is shown in Fig. 11(a), the BC and the DE parts of the profile denote the regions where the wall force field is dominant. Therefore, the summation of the thermal resistance of these two regions gives the total thermal resistance of the wall force field region in the gas as follows:

$$R_{Wall\ Force\ Filed} = R_{BC} + R_{DE} \qquad (6)$$

In addition to the above-mentioned definition, the CD part of the profile in the Fig. 11(a) is the bulk region where the wall force field loses its importance and argon atoms interact with each other freely. Therefore, the thermal resistance of the bulk region in the gas is defined as

$$R_{Bulk} = R_{CD}. \qquad (7)$$

In Table 2 we also show values obtained from equations 5-7 and the corresponding normalized values by $R_{Tot}$. This allows to understand the relative impact on the individual contributions. For example, if the interfacial thermal resistance forms a remarkable portion of the total resistance, changing the wall material would effectively change the heat transfer rate. Besides, it is expected that in the case where a considerable portion of total thermal resistance refers to the bulk region, changing the wall material would not change the total heat transfer rate considerably.

Furthermore, the relative importance of the wall force field resistance might be important in some cases, too. If for a specified case, it is negligible as compared to the total thermal resistance, then there would be no need for any molecular dynamics simulation of such a case, but a kinetic theory based method would effectively predict the heat transfer rate of the gas medium with lower computational cost. However, if the wall force field resistance forms a notable portion of the total resistance, the molecular dynamics method should be used.

The interfacial and total thermal resistance of the constant density and constant height case are shown in Figs. 11(b) and (c), respectively. Since the temperature difference between the walls is $\Delta T = 20$ K and the initial temperature of the gas $\left(T_{initial}^{gas}\right)$ is 298 K, $\Delta T \ll T_{initial}^{gas}$. Therefore, it is expected that the interfacial thermal resistance of the hot and the cold walls is of the same order (see Fig. 11(a)). Considering the constant height case, the gas density increases by reducing the Knudsen number. Therefore, more argon gas atoms are contributed to transfer heat to/from the walls as the Knudsen number is decreased. Therefore, a remarkable reduction in interfacial thermal resistance and total thermal resistance is expected as it is shown in Figs. 11(b) and (c), respectively. Actually, when the argon gas density is increased from 1.896 kg/m$^3$ to 189.6 kg/m$^3$, a reduction in the interfacial thermal resistance near the cold wall from 6.8 μKm$^2$/W to 0.075 μKm$^2$/W is observed while near the hot wall, the interfacial thermal resistance changes from 7.4 μKm$^2$/W to 0.075 μKm$^2$/W. At the same time, the total thermal resistance reduces from 26.15 μKm$^2$/W to 0.445 μKm$^2$/W. Thus, increasing the density from a rarefied to a dense gas condition drastically reduces the thermal resistance of the gas.

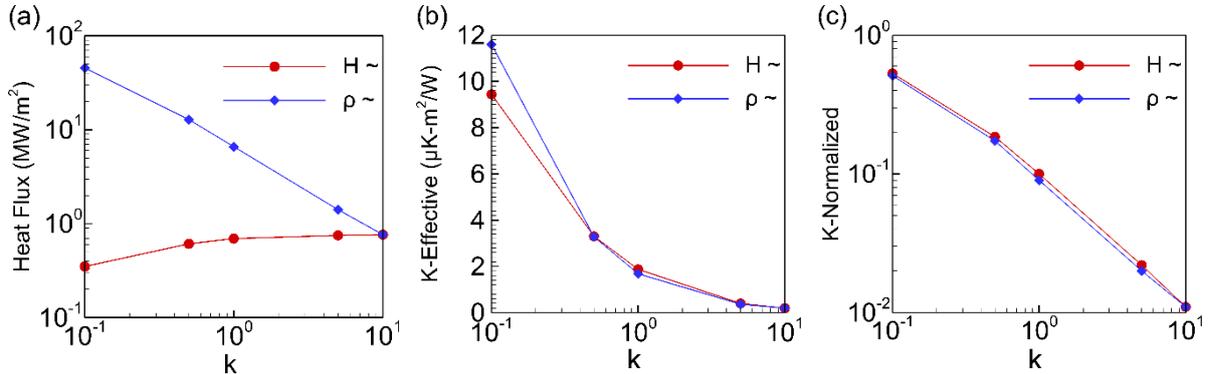

**Fig. 10.** Variation of heat flux **(a)**, effective thermal conductivity **(b)** and normalized effective thermal conductivity **(c)** of the argon gas for different modified Knudsen numbers

For the constant density case, the channel height increases by reducing the Knudsen number. According to Fig. 11(c), the total thermal resistance increases from 26.15 μKm²/W to 57.13 μKm²/W while the channel height increases from $H = 5.4$ nm to $H = 540$ nm. This increase in total thermal resistance refers to the fact that as the channel height increases, the temperature gradient between the walls decreases and consequently the thermal resistance of the gas increases. Considering the gas at $k = 10$, Table 2 shows that more than half of the total thermal resistance is due to the interface effect. Considering the fact that heat transfer at the interface is due to phonon dispersion between two dissimilar materials [42] it is obvious that due to the rarefaction of the argon gas, the phonon mismatch between the gas and solid walls is considerable in this case. Furthermore, it is clear from Table 2 that as the Knudsen number is decreased, the relative contribution of the interface thermal resistance in comparison to the total thermal resistance is decreased.

Fig. 11(b) clearly shows that in the constant density case, the interfacial thermal resistance is reduced with a small slope. The reason refers to the fact that in the constant density case, the argon gas density is kept constant and since the wall surface and temperature are the same for all Knudsen numbers, it is expected that the interfacial thermal resistance does not vary too much. Considering Table 2, $R_{Interfacial}$ reduces by about 20% from 13.94 μKm²/W to 11.13 μKm²/W when $k$ changes from 10 to 0.1. Since the gas density and wall temperature are the same for both cases, the reduction in interfacial thermal resistance is expected to change the heat transfer mechanism from ballistic for the $k = 10$ case to diffusive for $k = 0.1$.

**Table 2**
Thermal resistance ($\mu Km^2/W$) of the argon gas for the constant density and the constant height cases

| $k$ | $R_{AB}$ | $R_{BC}$ | $R_{CD}$ | $R_{DE}$ | $R_{EF}$ | $R_{Tot}$ | $\frac{R_{Interfacial}}{R_{Tot}}$ | $\frac{R_{Wall\ Force\ Filed}}{R_{Tot}}$ | $\frac{R_{Bulk}}{R_{Tot}}$ |
|---|---|---|---|---|---|---|---|---|---|
| 0.1 ($H\sim$) | 5.43 | 5.14 | 35.83 | 5.03 | 5.7 | 57.13 | 0.19 | 0.18 | 0.63 |
| 0.1 ($\rho\sim$) | 0.075 | 0.025 | 0.24 | 0.03 | 0.075 | 0.445 | 0.34 | 0.12 | 0.54 |
| 0.5 ($H\sim$) | 5.8 | 5.47 | 10.16 | 5.52 | 6.01 | 32.96 | 0.36 | 0.33 | 0.31 |
| 0.5 ($\rho\sim$) | 0.34 | 0.22 | 0.43 | 0.23 | 0.34 | 1.56 | 0.44 | 0.29 | 0.27 |
| 1 ($H\sim$) | 6.11 | 5.37 | 5.6 | 4.49 | 6.33 | 28.25 | 0.44 | 0.35 | 0.21 |
| 1 ($\rho\sim$) | 0.71 | 0.51 | 0.6 | 0.51 | 0.71 | 3.04 | 0.47 | 0.34 | 0.19 |
| 5 ($H\sim$) | 6.66 | 4.97 | 2.84 | 5.46 | 6.86 | 26.79 | 0.50 | 0.39 | 0.11 |
| 5 ($\rho\sim$) | 3.46 | 2.95 | 1.11 | 3.01 | 3.55 | 14.18 | 0.49 | 0.42 | 0.09 |
| 10 | 6.8 | 5.61 | 1.19 | 5.51 | 7.14 | 26.15 | 0.53 | 0.42 | 0.05 |

In fact, at $k = 10$, the molecular motion and as a consequence the heat transfer mechanism is mainly the ballistic transport. In such a situation, the probability of collisions between the argon gas atoms and the walls is much higher than the probability of collisions between the argon gas atoms. Considering the fact that the presence of the wall force field region near each wall increases the residence time of gas atoms in this region [15,16], it is obvious that this additional residence time in the wall force field region reduces the ballistic transport heat transfer since it does not let the argon atoms move freely toward the other wall. Actually, the argon gas atoms should escape from this force field region to reach to the other wall which takes time. Consequently, the heat transfer ability at the surface is decreased and the interfacial thermal resistance is increased. On the other hand, the heat transfer mechanism at $k = 0.1$ is diffusive transport. In such a situation, the probability of collisions of the argon gas atoms is much higher than the probability of collisions with the wall. Therefore, the heat is transferred between the walls by the collision of the argon gas atoms with each other. In this regard, the increase in residence time in the near wall region does not affect the heat transfer too much since heat is transferred to the neighboring gas atoms in the wall force field region and in a similar way it dissipates in the whole gas.

Actually, by diffusive transport heat transfer, there is no need for argon atoms to escape from the wall force field region to be able to transfer the heat so the thermal resistance at the interface reduces by about 20% in comparison with the ballistic transport case. As mentioned before, in this case for $k = 0.1$, the channel height is $H = 540$ nm. Considering the fact that the wall force field is approximately effective up to 1 nm from each wall, only 0.0037 of the height of the channel is affected by the wall force field. Interestingly, according to Table 2, this small portion of the channel's height is responsible for 0.18 of total thermal resistance. By increasing the argon gas atoms residence time in the wall force region causes an increased gas temperature in the vicinity of the walls. Even though the channel height is almost reaching microscale dimensions (540 nm), it is apparent that the wall force field and the interfacial thermal resistance are still forming a considerable portion of the total thermal resistance.

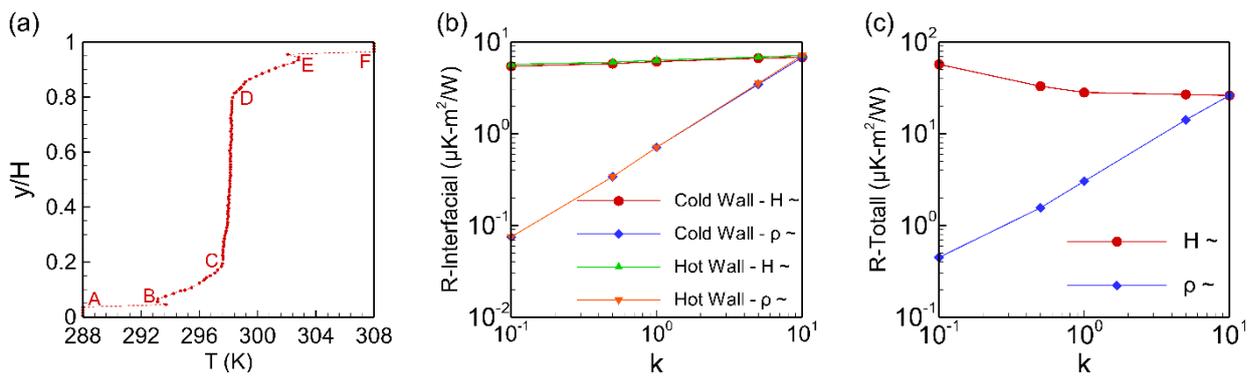

**Fig. 11.** Decomposition of the temperature profile into five regions **(a)**, interface thermal resistance **(b)** and total thermal resistance **(c)** of the argon gas for different Knudsen numbers

Sofos et al. [43] have shown that for liquid argon flow confined in the nanochannel, the thermal conductivity is higher in the bulk region of liquid medium compared to the layers adjacent to the walls. In order to investigate the effect of the wall force field on transport properties of the nano-confined gas, the gas medium is divided into three distinct regions (BC, CD and DE) where the temperature profile can be approximated linearly in each region, see Fig. 11(a). Therefore, we are able to use the effective thermal conductivity formula locally ($K_{eff} = J\,\Delta T/H$) for each region and define an effective local thermal conductivity ($K_{eff}^L$) for each region separately. The calculated values for $K_{eff}^L$ is shown in the Fig. 12(a) and Fig. 12(b) for $H\sim$ and $\rho\sim$ cases respectively. As it is shown in this figure, the wall force field reduces the thermal conductivity of the layers adjacent to the walls considerably in comparison with the bulk region value. It should be noticed that while the calculated values of $K_{eff}^L$ in the bulk region are in the same order for $H\sim$ and $\rho\sim$ cases, the corresponding values for the wall force field region strongly depends on the changing the Knudsen number mechanism. Figure 12(a) clearly shows that for the $H\sim$ case where the density is constant, $K_{eff}^L$ is in the order of 0.12 mW/mK regardless of the Knudsen number. On the other hand, for the $\rho\sim$ case where the density increased gradually, $K_{eff}^L$ varies from 0.12 mW/mK to 11 mW/mK as the Knudsen number changes from 10 to 0.1. These observations imply that the determinative factor for gas transport characteristics in wall force field region is the gas density. Actually, as the gas density is increased, more energy carrier units (gas atoms) are involved in heat transfer phenomena which leads to a higher $K_{eff}^L$.

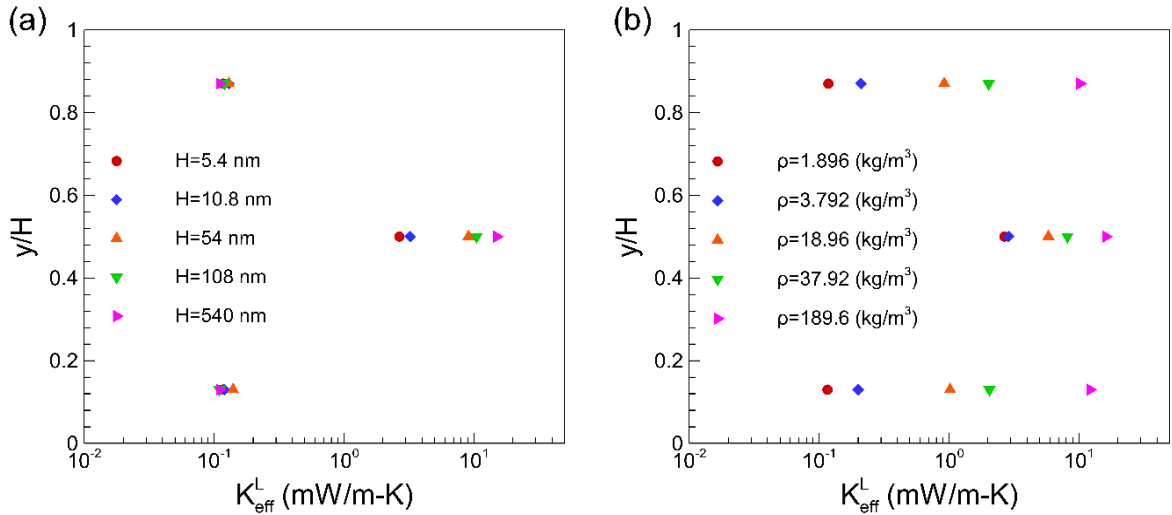

**Fig. 12.** Local thermal conductivity distribution for three different layer of gas along the channel height for the H~ case **(a)** and the ρ~ case **(b)**

## 5. Summary and conclusions

The effect of the wall force field on the distribution of various gas properties in the presence of heat transfer is investigated using three-dimensional molecular dynamics simulations of a stationary argon gas confined in a nanochannel. The top wall temperature is fixed at 308 K and the bottom wall is kept at 288 K. The significant importance of the wall force field on the distribution of temperature and the density profile in approximately 1 nm from each wall is observed for all Knudsen numbers in the transition regime regardless of the channel height and the gas density. Meanwhile, the maximum value for the temperature profile and the normalized density distribution can be found at $\sigma/2$ from each wall for all Knudsen numbers.

It is observed that for the constant density case, in which the channel height is increased gradually from 5.4 nm to 540 nm while $\rho = 1.896$ kg/m$^3$, the normalized density distribution in the wall force field region is independent of the Knudsen number and it is only related to the wall temperature. Unlike this case, for constant height case in which the channel height is $H = 5.4$ nm and the density changes from 1.896 kg/m$^3$ to 189.6 kg/m$^3$, the maximum value of the normalized gas density is decreased gradually in the wall force field region. This difference leads to a reduction of the temperature jump for the constant density case in comparison with the constant height case. The temperature profile along the channel height shows a similar behavior regardless of the method used to change the Knudsen number. While the wall force field changes the temperature distribution within approximately 2 nm from each wall and reduces the temperature jump between the gas and the surface, in the bulk region it changes from an approximately constant value for $k = 10$ toward a linear variation between the wall's temperatures for $k = 0.1$. The variation of effective thermal conductivity as an indicator of heat transfer ability of a medium shows a nonlinear increase as the Knudsen number is decreased for both methods of changing the Knudsen number. Interestingly, the normalized effective thermal conductivities of both cases coincide with each other.

In addition, for $k = 10$, 0.53 of the total thermal resistance is due to the interfacial thermal resistance and 0.42 of the total thermal resistance arises from wall force field regions, while the bulk region resistance is negligible. This shows that the gas atoms are highly affected by wall force field region. It is shown that reducing the Knudsen number in the constant height case leads to a reduction in the interfacial, wall force field and total thermal resistance. In contrast, decreasing the Knudsen number in the constant height case leads to an increase in the total thermal resistance while a small reduction of about 20% in interfacial and wall force field thermal resistance is observed due to the change in heat transfer mechanism from ballistic in the free molecular regime to diffusive in the near continuum limit. It is also interesting to note that the wall force fields along with the interface thermal resistance are forming about 37% of the total thermal resistance even for $H = 540$ nm. This shows the importance of considering these two thermal resistances for the calculation of the total thermal resistance in a wide range of the channel height (5.4 nm to 540 nm).

Furthermore, it is observed that the local thermal conductivity is reduced significantly in comparison to the bulk value which is referred to the presence of the wall force field. While the local thermal conductivity in the bulk region is observed to be approximately independent of changing the Knudsen number mechanism, for the near-wall regions, the local thermal conductivity strongly depends on it. It is shown that by increasing the gas density to change the Knudsen number, the local thermal conductivity in the wall force field region increases as well. In contrast, if the Knudsen number is tuned by changing the channel height, the local thermal conductivity in the wall force field region stays approximately constant.

## Acknowledgment

This research was supported financially by the research council of the Tarbiat Modares University through a Ph.D. student fellowship to Reza Rabani.
## References

[1]     J. Li, Y. Lu, Q. Ye, M. Cinke, J. Han, M. Meyyappan, Carbon nanotube sensors for gas and organic vapor detection, Nano Lett. 3 (2003) 929–933. doi:10.1021/nl034220x.

[2]     I.K. Hsu, M.T. Pettes, M. Aykol, L. Shi, S.B. Cronin, The effect of gas environment on electrical heating in suspended carbon nanotubes, J. Appl. Phys. 108 (2010) 84307. doi:10.1063/1.3499256.

[3]     J.-Y. Juang, D.B. Bogy, C.S. Bhatia, Design and Dynamics of Flying Height Control Slider With Piezoelectric Nanoactuator in Hard Disk Drives, J. Tribol. 129 (2007) 161. doi:10.1115/1.2401208.

[4]     M. Kurita, Junguo Xu, M. Tokuyama, K. Nakamoto, S. Saegusa, Y. Maruyama, Flying-height reduction of magnetic-head slider due to thermal protrusion, IEEE Trans. Magn. 41 (2005) 3007–3009. doi:10.1109/TMAG.2005.855240.

[5]     H. Li, C.T. Yin, F.E. Talke, Thermal insulator design for optimizing the efficiency of thermal flying height control sliders, J. Appl. Phys. 105 (2009) 07C122. doi:10.1063/1.3074558.

[6]     G.A. Bird, Molecular gas dynamics and the direct simulation of gas flows, Clarendon Press, Oxford, UK, 1994.

[7]     C. Cercignani, The Boltzmann Equation and Its Applications, Springer New York, New York, NY, 1988. doi:10.1007/978-1-4612-1039-9.

[8]     G. Karniadakis, A. Beskok, A. Narayan, Microflows and Nanoflows, Springer, 2005. doi:10.1007/0-387-28676-4.

[9]     J. Toth, Adsorption : theory, modeling, and analysis, Marcel Dekker, New York, 2002.

[10]    L. Zhou, Adsorption - Progress in fundamental and application research, World Scientific, 2007. doi:10.2533/chimia.2009.279.
19